\documentclass[conference, a4paper]{IEEEtran}
\IEEEoverridecommandlockouts
\usepackage{cite}
\usepackage{amsmath,amssymb,amsfonts,mathtools}
\usepackage{algorithmic}
\usepackage{graphicx}
\usepackage{textcomp}
\usepackage{svg}
\usepackage{xcolor}
\usepackage{multirow}
\usepackage{pifont}
\usepackage[hyperfootnotes=false, colorlinks=false, pdfborder={0 0 0}]{hyperref}
\usepackage[capitalise]{cleveref}
\usepackage{orcidlink} 
\usepackage{flushend} 
\usepackage[binary-units, detect-all=true]{siunitx}
\usepackage[printonlyused,nolist]{acronym}
\def\BibTeX{{\rm B\kern-.05em{\sc i\kern-.025em b}\kern-.08em
    T\kern-.1667em\lower.7ex\hbox{E}\kern-.125emX}}

\usepackage[all]{background}
\usepackage{stackengine}
\setstackEOL{\\}
\setstackgap{L}{\normalbaselineskip}
\SetBgContents{\color{gray}{\tiny \Longstack{PREPRINT - accepted by 31th IFIP/IEEE International Conference on Very Large Scale Integration 2023 (VLSI-SoC 2023)}}}
\SetBgPosition{3.5cm,1cm}
\SetBgOpacity{1.0}
\SetBgAngle{0}
\SetBgScale{1.8}

\begin{document}

\bstctlcite{IEEEexample:BSTcontrol}

\begin{acronym}
    \acro{rram}[RRAM]{resistive random access memory}
    \acro{cim}[CIM]{computing-in-memory}
    \acro{mvm}[MVM]{matrix-vector multiplication}
    \acro{sram}[SRAM]{static random-access memory}
    \acro{cnn}[CNN]{convolutional neural network}
    \acro{noc}[NoC]{Network-on-Chip}
    \acro{isa}[ISA]{instruction set architecture}
    \acro{vg}[VG]{vertical group}
    \acro{hg}[HG]{horizontal group}
    \acro{ifm}[IFM]{input feature map}
    \acro{ofm}[OFM]{output feature map}
    \acro{mvm}[MVM]{matrix-vector multiplication}
    \acro{gpu}[GPU]{graphics processing unit}
    \acro{tpu}[TPU]{tensor processing unit}
    \acro{gpeu}[GPEU]{general purpose execution unit}
\end{acronym}

\def\finalpaper{1} 
\title{Mapping of CNNs on multi-core RRAM-based CIM architectures}
\if\finalpaper1
    \author{
        \IEEEauthorblockN{Rebecca Pelke\orcidlink{0000-0001-5156-7072}, Nils Bosbach\orcidlink{0000-0002-2284-949X}, Jose Cubero\orcidlink{0000-0001-9575-0856}, Felix Staudigl\orcidlink{0000-0001-9673-3070},
            Rainer Leupers\orcidlink{0000-0002-6735-3033}, and Jan Moritz Joseph\orcidlink{0000-0001-8669-1225}}
        \IEEEauthorblockA{
            \textit{RWTH Aachen University, Germany}\\
        }
        \thanks{This work was funded by the Federal Ministry of Education and Research (BMBF, Germany) in the project NeuroSys (Project Nos. 03ZU1106CA).}
        \vspace{-1cm}
    }
\else
    \author{
      \IEEEauthorblockN{Authors are removed for submission version}
      \\
      \\
      \IEEEauthorblockA{Affiliations are removed for submission version}
      \vspace{-1cm}
      \\
    }
\fi

\IEEEoverridecommandlockouts
\IEEEpubid{\makebox[\columnwidth]{979-8-3503-2599-7/23/\$31.00~\copyright2023 IEEE\hfill}
\hspace{\columnsep}\makebox[\columnwidth]{ }}

\maketitle

\begin{abstract}
\Ac{rram}-based multi-core systems improve the energy efficiency and performance of \acp{cnn}.
Thereby, the distributed parallel execution of convolutional layers causes critical data dependencies that limit the potential speedup.
This paper presents synchronization techniques
for parallel inference of convolutional layers on
\ac{rram}-based \ac{cim} architectures.
We propose an architecture optimization that enables efficient data exchange 
and discuss the impact of different architecture setups on the performance.
The corresponding compiler algorithms are optimized for high speedup and low memory consumption during \ac{cnn} inference.
We achieve more than \SI{99}{\percent} of the theoretical
acceleration limit with a marginal data transmission overhead of
less than \SI{4}{\percent} for state-of-the-art \ac{cnn} benchmarks.

\end{abstract}
\vspace{-0.25cm}

\section{Introduction}
\acresetall 

In recent years, the broad use of \acp{cnn} in computer vision applications yielded an ever-growing demand for efficient architectures to handle these compute- and data-intensive workloads.
Due to the massive parallelism and reuse capabilities in \acp{cnn}, these applications are not only executed on classical von-Neumann architectures but also on specialized hardware including \acp{gpu} and \acp{tpu}.
Today, \ac{cnn} accelerators exist in various form factors, from power-efficient edge devices to hyper-scaled compute clusters.

Despite the extensive innovation sparked by the ubiquitous use of \acp{cnn}, all these custom architectures suffer from one major performance limitation, namely, moving data from the system's main memory to the compute units, and vice versa~\cite{jouppi2017datacenter}.
In other words, \ac{cnn} accelerators suffer from the von-Neumann bottleneck~\cite{zou2021breaking}.
Novel \ac{cim} technologies, such as \ac{rram}, promise to tackle this bottleneck by unifying memory and computation unit~\cite{chang2017memcomputing}.
These designs offer a significant advantage over CMOS-based designs in terms of memory capacity, device density, and power consumption~\cite{vetter2015opportunities}.

Previous works presented accelerator architectures that use \ac{rram} crossbars as \ac{mvm} units~\cite{NeuRRAM, ankit2019puma, shafiee2016isaac, chi2016prime}.
These architectures are designed hierarchically to scale from single \ac{mvm} units to complex multi-core systems.
To achieve maximum flexibility and scalability, the \ac{mvm} units are often embedded in \ac{cim} cores, which can communicate with each other over a bus system~\cite{ankit2019puma, shafiee2016isaac}.

The accelerators aim at a weight stationary data flow, i.e., the weights of the \ac{cnn} are statically assigned to \ac{rram} crossbars \cite{liu2021fpra}.
This requires the development of new concepts in the compiler domain.
The authors of \cite{yanai2016efficient,zhang2020efficient,rhe2022vw,rhe2022vwc,negi2022nax,agrawal2019x} investigated the translation of conv2D operations to \acp{mvm}.
They focused on the mapping of kernel weights to \ac{rram} crossbars.
To achieve a high speedup, the kernel weights of one layer must be split across multiple \ac{cim} cores for parallel processing.
This causes critical data dependencies between cores, which are often neglected.
Synchronization techniques are needed to resolve these dependencies.
They must be considered in the context of the underlying architecture.
The authors of \cite{ankit2019puma} proposed a centrally organized synchronization technique.
This scheme requires a high amount of non-\ac{rram} memory.

\begin{figure}[tbp]
    \centering
    \includegraphics[width=.8\linewidth, keepaspectratio]{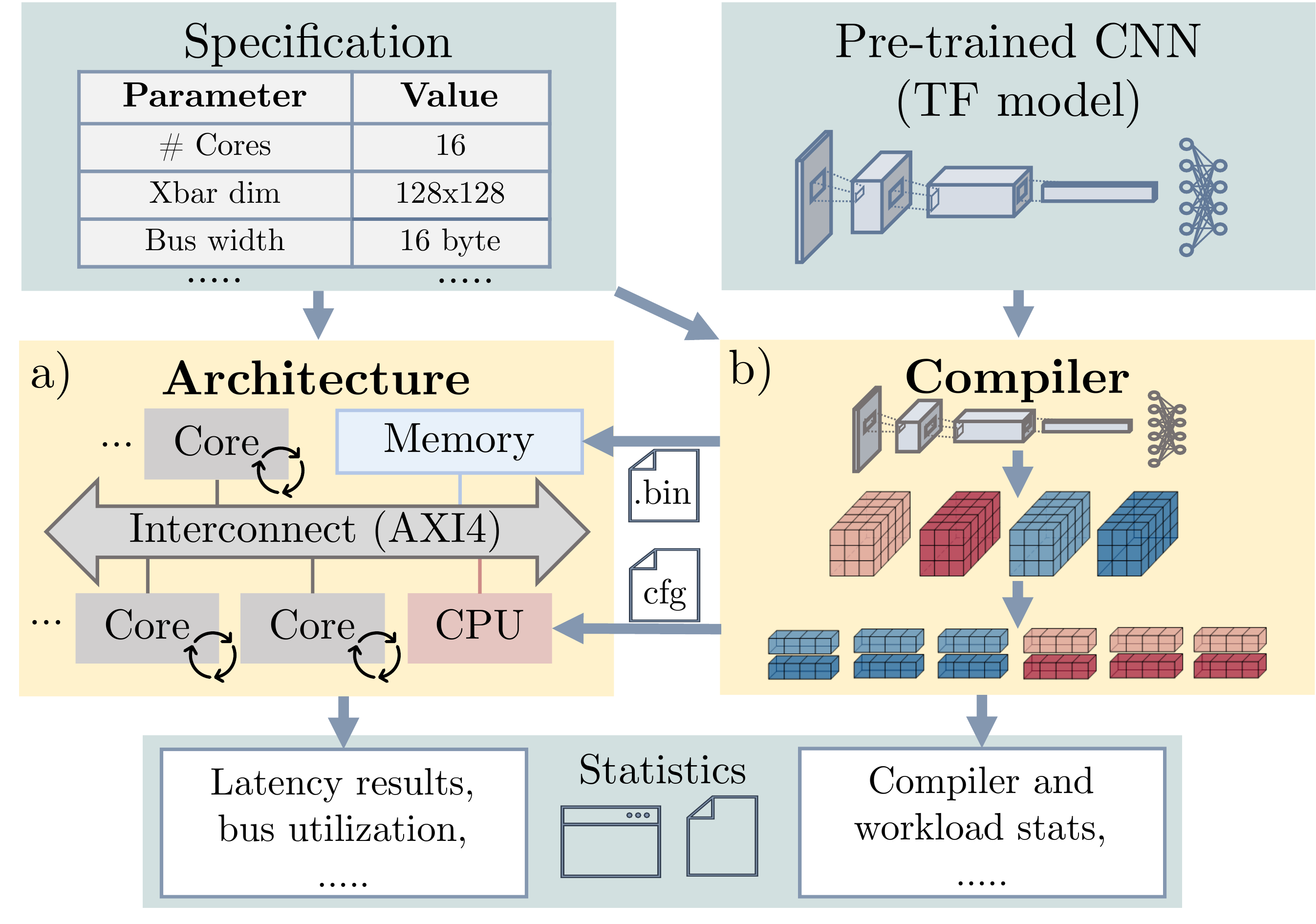}
    \caption{Evaluation framework containing architecture (a) and compiler (b).}
    \label{fig:overview}
    \vspace{-10pt}
\end{figure}

In our work, we enable efficient, low-overhead parallel execution of \ac{cnn} layers on multi-core \ac{rram}-based \ac{cim} architectures.
We use decentralized synchronization methods to minimize memory consumption with marginal data traffic overhead.
This includes the following contributions:
\begin{itemize}
  \item An architecture optimized for efficient, decentralized, and event-based communication.
  \item Compiler algorithms that achieve more than \SI{99}{\percent} of the theoretical acceleration limit for conv2D layers.
  \item A cycle-accurate simulator to analyze the influence of different algorithms and architecture parameters.
\end{itemize}

\cref{fig:overview} illustrates our evaluation framework.
The architecture simulator is used to validate and evaluate the proposed algorithms (see \cref{fig:overview}(a)).
The specification allows setting different architecture parameters to investigate their influence on the \ac{cnn} inference.
The compiler receives a \ac{cnn} model and an architecture specification as input and generates code that can be executed on the simulator
(see \cref{fig:overview}(b)).

\section{Background}

\subsection{\acs{rram} crossbars}
A \ac{rram} device is a non-volatile emerging memory that stores a conductance value. Multiple \ac{rram} devices can be arranged in crossbar structures to enable in-memory computing~\cite{cao2021neural}.
On \ac{rram} crossbars, \acp{mvm} can be performed in the analog domain in $\mathcal{O}(1)$~\cite{li2015rram}.
The weights of the neural network are stored in the crossbar.
By applying the input values as voltages, currents are generated that correspond to the result of the \ac{mvm}.
Modern \ac{rram} crossbars have been fabricated in different sizes, e.g.,
$64\times64$\cite{ankit2019puma}, $128\times128$\cite{shafiee2016isaac}.

\subsection{Weight mapping}
Performing \acp{mvm} is significantly faster than programming the crossbar cells \cite{NeuRRAM}.
If the accelerator provides a sufficient number of crossbars, the weights should therefore be programmed only once to ensure an efficient inference phase.
Conv2D layers are well suited for the execution on \ac{rram} crossbars since they can be translated into \acp{mvm} and the matrices can be reused multiple times.
This has been investigated in previous works.
One of the first methods, im2col~\cite{yanai2016efficient}, assigns kernel values to crossbar cells with the densest \ac{rram} cell occupancy.
In other approaches, the crossbar is more sparsely packed and kernel values are duplicated to increase the input reuse~\cite{zhang2020efficient, rhe2022vwc}.
Since the im2col scheme requires the least number of \ac{rram} cells, an extended im2col method is used in this work, which splits the kernel values of one layer over several crossbars~\cite{negi2022nax, agrawal2019x}.

\subsection{RRAM-based \acs{cim} architectures}
Several \ac{rram}-based \ac{cim} architectures have been presented~\cite{ankit2019puma, chi2016prime, shafiee2016isaac, song2017pipelayer}.
They aim at efficient and parallel execution of \acp{mvm}.
Besides different interconnect models, they mainly differ in how autonomously the \ac{cim} cores can operate.
It ranges from simple \ac{mvm} units to powerful \ac{isa}-based cores \cite{mittal2018survey}.
Simple \ac{mvm} units can be driven and synchronized by a central control unit.
Autonomous cores, on the other hand, can execute workloads independently and do not need to be actively controlled.
This makes them more flexible and performant.
They require a more advanced synchronization procedure, which is addressed in this work.

\subsection{Synchronization techniques}
\label{sec:sync}
The parallel execution of layers causes critical data dependencies that can lead to incorrect results (see \cref{cha:operation_remapping}). 
This can be avoided at the cost of performance loss by executing the critical sections in sequence \cite{rhe2022vwc} (see \cref{sec:sync_schemes}).
For parallel execution, synchronization methods are required that need to be supported by the architecture.

The authors of \cite{ankit2019puma} introduced a central synchronization scheme.
In their architecture, several \textit{tiles} form the accelerator.
A \textit{tile} is structured similar to the architecture in \cref{fig:overview}(a) and contains a controller,
several \ac{cim} cores, and shared memory.
The synchronization is solved centrally by extending the shared memory with an attribute buffer.
This attribute buffer contains two attributes for each data entry, \textit{valid} and \textit{count}.
A memory controller maintains the attributes to ensure the correct exchange of data.
In this solution, a significant amount of memory is needed to store the attributes.
For \SI{64}{\kilo\byte} of data in the shared memory, \SI{32}{K} attributes are required~\cite{ankit2019puma}.
We improve on this idea by proposing a decentralized synchronization scheme that requires significantly less memory.

\section{Architecture}

We model a multi-core system as a reference architecture (see \cref{fig:overview}(a)).
The \ac{cim} cores are connected by a bus and use shared memory to exchange their data.
In this work, the bus system is used to execute one layer.
To be able to execute whole \acp{cnn}, the system can simply be duplicated.
Architecture parameters, e.g., the number of cores and the size of the crossbars, can be specified variably in our model to investigate their influence on the performance (see \cref{sec:results_arch_exploration}).

\begin{figure}[!tbp]
  \centering
  \includegraphics[width=.9\linewidth, keepaspectratio]{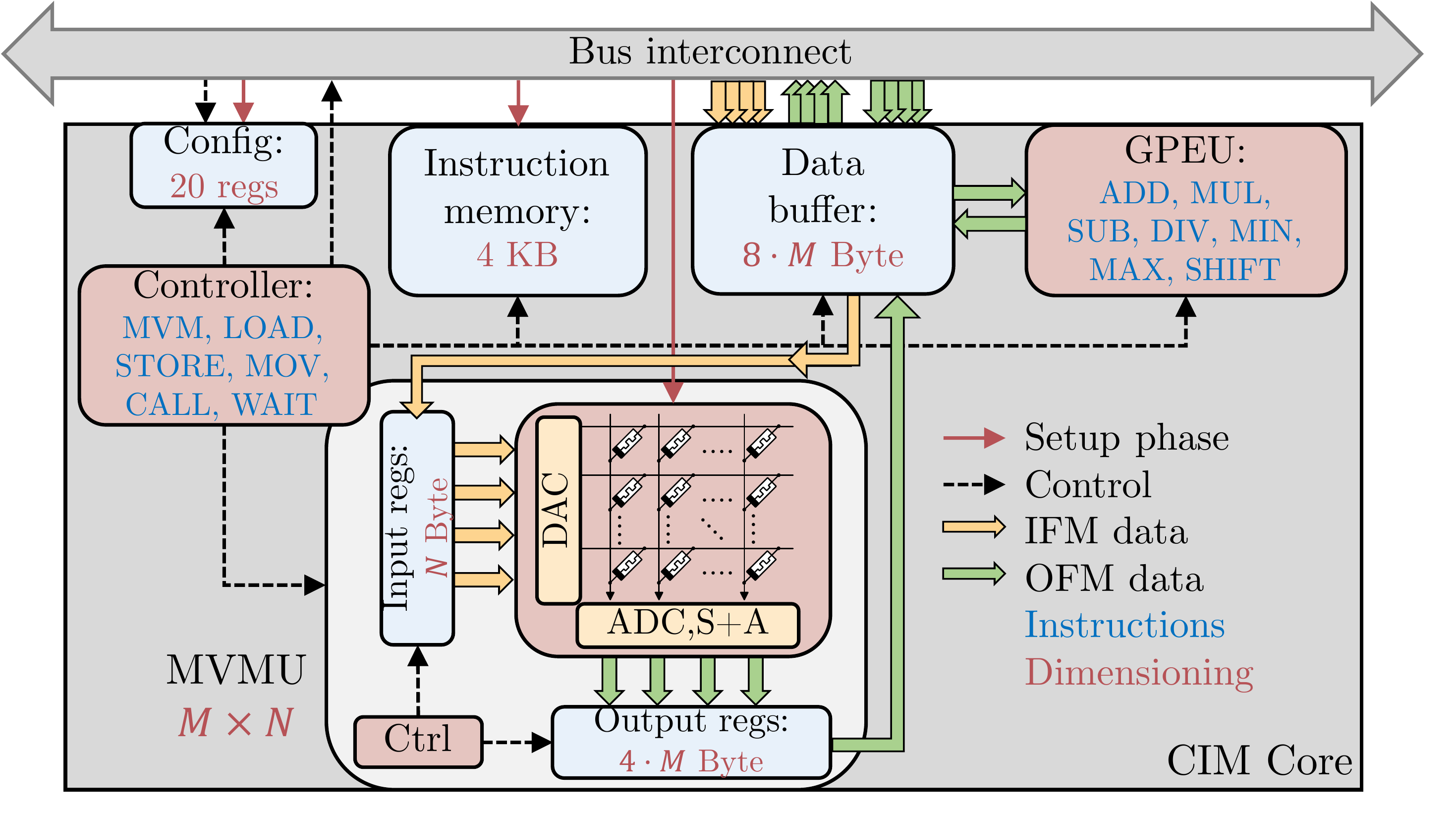}
  \caption{\ac{cim} core architecture, data flow, instructions, and dimensioning.}
  \label{fig:core}
  \vspace{-0.3cm}
\end{figure}

\cref{fig:core} illustrates the \ac{cim} core model including data flow, instructions, and dimensioning.
The buffer sizes depend on the matrix size that the \ac{mvm} unit can process.
Cores act as initiators and targets of bus transactions.
As initiators, they can, e.g., perform LOAD and STORE operations.
As a target, they can receive \textit{config} parameters.
The \ac{gpeu} can perform arithmetic operations and some activation functions like ReLU and LeakyReLU.

Cores operate in two phases.
In the setup phase, the CPU configures the \ac{cim} cores.
The instructions are loaded and kernel values are programmed into crossbar cells.
In the inference phase, the \ac{cnn} layer is executed.
After all calculations are completed, an interrupt is signaled to the CPU.

\section{Compiler}

Our compiler is written in Python to enable simple proofs of concept.
It compiles conv2D and dense layers from Tensorflow models and generates a \textit{bin} and a \textit{cfg} file for each layer depending on the architecture specification.
The \textit{cfg} file is interpreted by the CPU to configure the \ac{cim} cores in the setup phase.
The \textit{bin} file is loaded into the shared memory of the \ac{cim} cores.
It contains an instructions section for each core separately in case not all instructions fit into the instruction memory of the core.
It provides placeholders for the \ac{ifm} and \ac{ofm} of the layer.
In the following, mapping and synchronization techniques are discussed for the conv2D operation.
Dense layers can be treated analogously.

\subsection{Operation remapping}
\label{cha:operation_remapping}
\begin{figure}[!tbp]
  \centering
  \includegraphics[width=\linewidth, keepaspectratio]{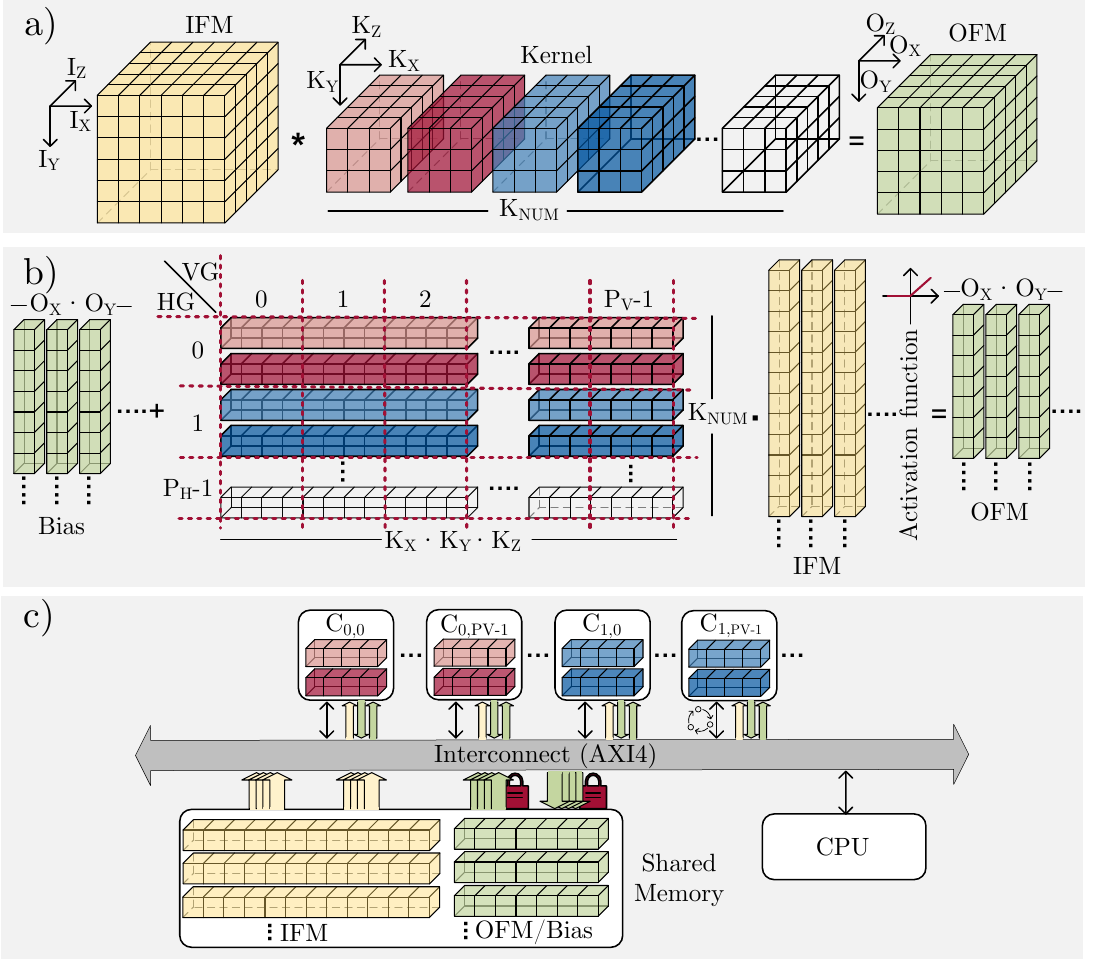}
  \caption{Translation of a conv2D operation (a) into multiple \acp{mvm} with matrix dimension $M\times N$ (b) and distribution of kernel matrix values to cores (c).}
  \label{fig:remapping}
  \vspace{-0.3cm}
\end{figure}

\cref{fig:remapping}(a) shows the main components of a conv2D layer, i.e., \ac{ifm}, \ac{ofm}, and kernels.
\cref{fig:remapping}(b) illustrates the im2col scheme~\cite{agrawal2019x}. 
The unrolled kernels form a matrix, which can then be multiplied by $O_X \cdot O_Y$ unrolled vectors from the \ac{ifm}.

State-of-the-art \ac{cnn} layers are often too big to be stored in a single crossbar.
The kernel values of one layer have to be split over several crossbars (red lines)~\cite{yanai2016efficient, negi2022nax}.
\cref{fig:remapping}(c) shows the assignment of kernel values to cores.
In the setup phase, the CPU loads the \ac{ifm} to the associated placeholder in the shared memory.
Bias values are initially written to the placeholder of the \ac{ofm}.
The \ac{gpeu} is used to accumulate the bias values and to accomplish the activation function.

We extend the multi-core im2col scheme by assigning two group IDs to each core. All cores sharing the same \ac{vg} ID operate on the same values of the \ac{ifm}. All cores sharing the same \ac{hg} ID generate partial results for the same values in the \ac{ofm} that have to be accumulated.
In the following, the cores are denoted as $C_{HG,VG}$.
While the \ac{ifm} is read-only, both, read and write accesses are performed on the \ac{ofm}. 
Considering $M\times N$ crossbars and $(K_Y, K_X, K_Z, K_{NUM})$ conv2D kernels (HWIO layout), the total number of needed cores $C_{NUM}$ is
\begin{equation*}
    C_{NUM} = \underbrace{\left\lceil \frac{K_X \cdot K_Y  \cdot K_Z}{N} \right\rceil}_{\eqqcolon P_{V}} \cdot
    \underbrace{\left\lceil \frac{K_{NUM}}{M} \right\rceil}_{\eqqcolon P_{H}}.
\end{equation*}

The area in the shared memory dedicated to the \ac{ofm} is reused for the exchange of partial results.
This keeps the \ac{cim} cores lean since the required buffer sizes and synchronization complexity remain minimal.
As a consequence, all cores sharing the same \ac{hg} ID operate on the same \ac{ofm} locations in the shared memory.
The access must be regulated to avoid race conditions.
Hence, a synchronization technique is required to ensure that all partial results are accumulated correctly.
This means $P_{H}$ sets of $P_{V}$ cores need to be synchronized for $O_{V,NUM} = O_X \cdot O_Y$
different output vectors of size $M$.

\subsection{Synchronization schemes}
\label{sec:sync_schemes}
\begin{figure*}[!tbp]
  \centering
  \includegraphics[width=.9\linewidth, keepaspectratio]{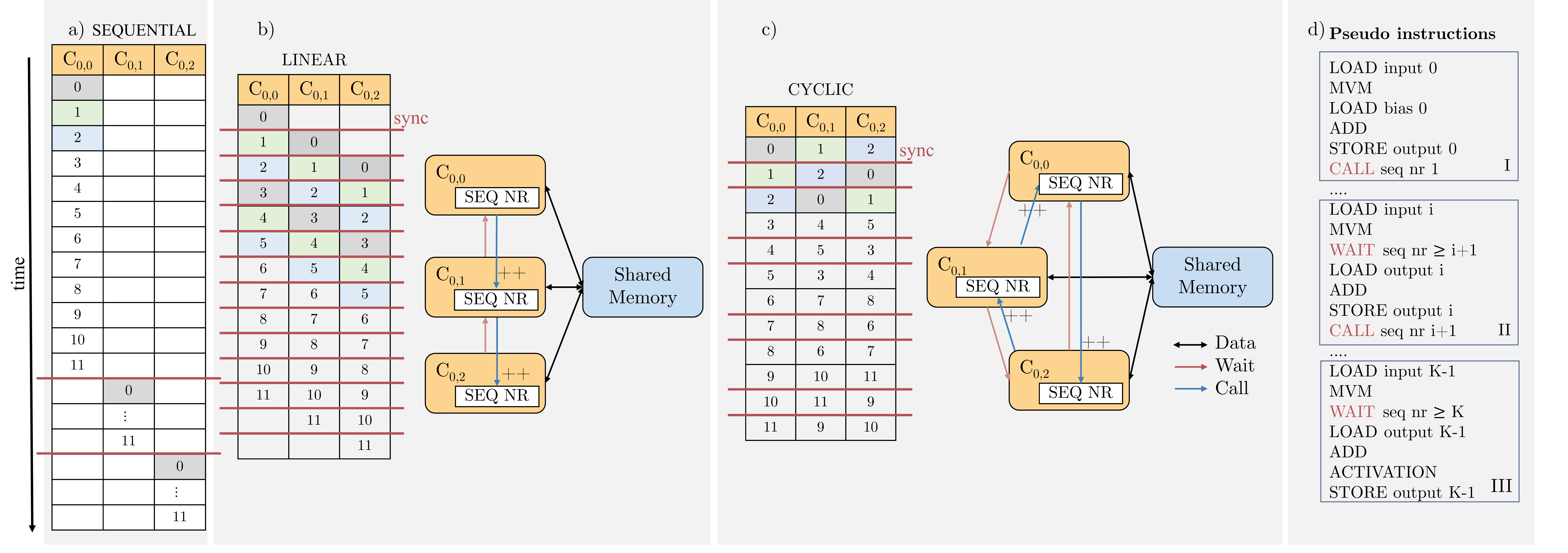}
  \caption{Sequential computation of \ac{ofm} without synchronization scheme (a), parallel computation of \ac{ofm} with linear (b) and cyclic (c) synchronization for conflicting cores $C_{0,0}, C_{0,1}$ and $C_{0,2}$, pseudo instructions for parallel computation of a conv2D operation (d).}
  \label{fig:sync}
  \vspace{-0.3cm}
\end{figure*}

All output vectors of the \ac{ofm} stored in the shared memory can be treated as a resource that may only be owned by one core at any time.
\cref{fig:sync} illustrates the proposed synchronization techniques for the example of three conflicting cores $C_{0,0}, C_{0,1}$, and $C_{0,P_V-1}=C_{0,2}$ ($P_{H}=1, P_{V}=3$) with $12$ different output vectors, i.e., $12$ different resources.
To calculate correctly, each core must have owned each resource once to accumulate its partial results.

In the following, the different parallelization and synchronization schemes are presented.
A red \textit{sync} line means that the core that releases a resource notifies (CALL) its successor. That is the core that will receive the resource next.
The successor must wait (WAIT) for the notification.

\subsubsection{Sequential synchronization}
This scheme is the most basic one. 
It is used in \cite{rhe2022vw,rhe2022vwc}.
Conflicting cores operate sequentially and not in parallel, which eliminates the need for complex synchronization procedures.
In the example, $C_{0,0}$ gets all resources first.
After it has completed all calculations, the next core, $C_{0,1}$, is allowed to operate.
The first core, $C_{0,0}$, accumulates the bias values and the last core, $C_{0,P_V-1}$, applies the activation function to all output vectors (see \cref{fig:sync}(a)).

In the following, we propose two schemes, \textit{linear} and \textit{cyclic} synchronization, to achieve parallel processing, i.e., cores of the same \ac{hg} can operate in parallel.

\subsubsection{Linear synchronization}
The cores process the output vectors in the same order, 
starting from $C_{0,0}$ to $C_{0,2}$ in the example.
Core $C_{0,0}$ has no predecessor and $C_{0,2}$ has no successor.
Core $C_{0,0}$ accumulates the bias values and $C_{0,P_V-1}$ applies the activation function for all output vectors (see \cref{fig:sync}(b)).
In this case, the number of CALL and WAIT operations is
\vspace{-0.1cm}
\begin{equation*}
    P_{H} \cdot O_{V,NUM} \cdot \left ( P_{V} - 1 \right ).
\end{equation*}

\subsubsection{Cyclic synchronization}
In cyclic synchronization, tasks are distributed as fair as possible among the cores.
Each core has exactly one predecessor and one successor.
The output vectors are processed cyclically by the cores.
The core that first gains access to an output vector accumulates the bias values to its partial results.
The core that receives access to an output vector last executes the activation function (see \cref{fig:sync}(c)).
As a result, the execution of the activation functions and the addition of the bias values are shared equally among all cores.
In this case, the number of CALL and WAIT operations is
\vspace{-0.1cm}
\begin{equation*}
    P_{H} \cdot \left\lceil \frac{O_{V,NUM}}{P_{V}} \right\rceil \cdot P_{V} \cdot \left ( P_{V} - 1 \right ).
\end{equation*}

\subsection{Sequence number}
\label{cha:seq_number}
We extend each core with a single register to enable parallelization and synchronization on the architecture side.
That is illustrated in \cref{fig:sync}(b) and \cref{fig:sync}(c).
This register contains a sequence number (SEQ\_NR) which is writable for other cores.
The initial value is $0$.
A CALL operation increments the sequence number of the successor core (blue line).
When executing a WAIT operation, the core waits for its sequence number to reach at least a certain value.
\cref{fig:sync}(d) shows the pseudo instructions.
Three cases are distinguished.
In the first case, the core has no predecessor for the output it is working on.
In the second case, the core has both, a predecessor and a successor.
The last case describes the scenario in which a core is the last one operating on an output.

\section{Results}

We evaluate our proposed parallelization techniques in terms of speedup gain and overhead costs using the conv2D layers of Mobilenet~\cite{howard2017mobilenets} and ResNet-18\cite{he2016resnet}.
Those layers are also found in other \acp{cnn}.
The synchronization methods do not affect the accuracy of the \acp{cnn}, which is therefore not examined here.
The conv2D layers are compiled for different architecture parameters and synchronization schemes to investigate their effects on the performance.

\subsection{Architecture simulator}
Compiled layers are executed on our SystemC/TLM-2.0-based simulator to verify the compiler concepts and algorithms. 
To obtain realistic latency values and enable architectural exploration on a high abstraction level, the TLM-2.0 non-blocking interface is used in combination with the approximately-timed coding style~\cite{aynsley2009osci}.
We use a multi-initiator-multi-target AXI4 bus interconnect~\cite{arm-dev-guide}.
The AXI4 bus protocol~\cite{axi-spec} supports burst transactions and out-of-order transaction completion, which are beneficial features for data-intensive and highly parallel workloads.
We capture relevant data during runtime to evaluate the impact of different architecture parameters and synchronization methods on the performance \cite{bosbach2022nistt}.

\subsection{Performance speedup}%
\begin{figure*}[!tbp]
  \centering
  \includegraphics[width=\linewidth, keepaspectratio]{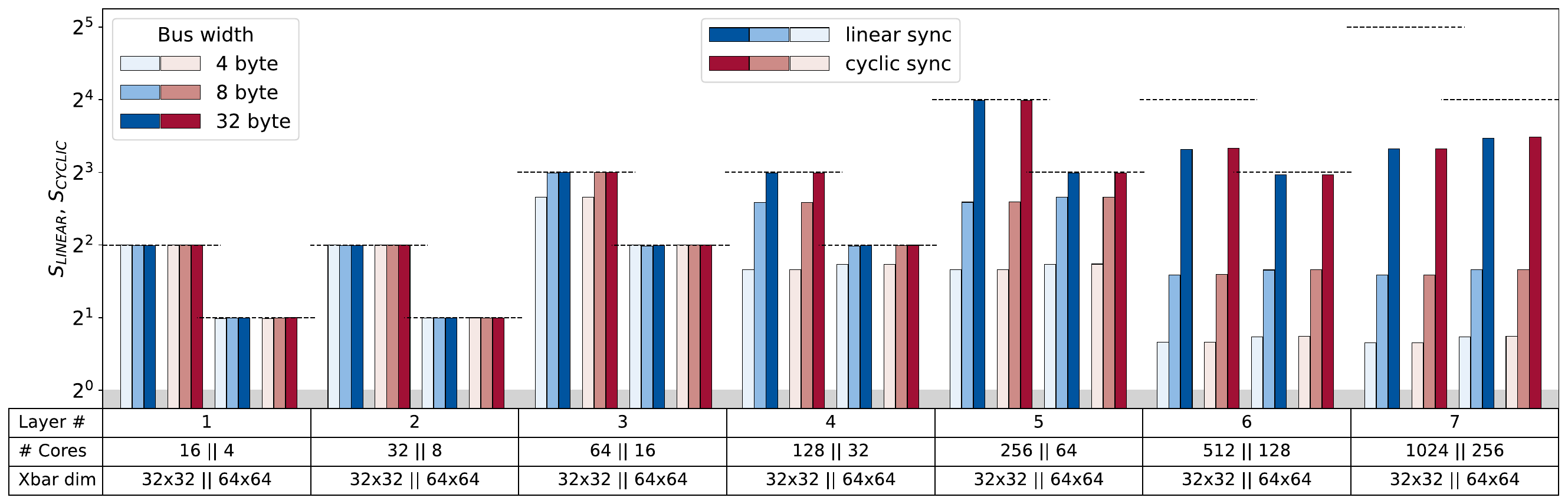}
  \caption{Speedup vs. maximum achievable speedup (dashed lines)
  of the linear and cyclic synchronization for different layers, crossbar dimensions, and bus widths.
  The number of cores depends on the layer and crossbar dimension.
  It refers to $32\times 32$ crossbars (left entry) and $64\times 64$ crossbars (right entry).}
  \label{fig:results_bar}
  \vspace{-0.4cm}
\end{figure*}
To evaluate the parallelization and synchronization methods, we examine the speedup of the linear (\cref{fig:sync}(b)) and cyclic schemes (\cref{fig:sync}(c)).
The speedup always refers to the latency of the corresponding sequential version \cite{rhe2022vw,rhe2022vwc} (\cref{fig:sync}(a)).
\begin{equation*}
    S_{LINEAR} = \frac{t_{SEQUENTIAL}}{t_{LINEAR}}, S_{CYCLIC} = \frac{t_{SEQUENTIAL}}{t_{CYCLIC}}
\end{equation*}
The variable $t_{X}$ denotes the latency of the inference of a conv2D layer using scheme $X$.
An upper bound for the maximum achievable speedup is $P_{V}$, i.e., all conflicting cores run in parallel without synchronization overhead (see \cref{sec:sync_schemes}).

\cref{fig:results_bar} shows the speedup of the linear (blue) and the cyclic synchronization method (red) for different conv2D layers of Mobilenet.
The shapes of the used layers (Layer \#) are listed in \cref{tab:mobilenet}.
The crossbar dimensions are $32\times 32$ and $64\times 64$. 
This, in combination with the kernel shape of the layer, determines the number of needed cores.
The upper bound for the maximum achievable speedup ($P_{V}$) is indicated by the dashed lines.
The speedup increases when reducing the crossbar size.
A reduction from $64\times 64$ to $32\times 32$ crossbars results in a speedup of at most $2\times$ referred to the corresponding sequential scheme.
Up to $4\times$ more cores are required which increases the bus utilization and synchronization complexity.
This means that higher speedups can be achieved at the cost of higher numbers of cores and larger bus widths.

\cref{fig:results_bar} also reveals the speedup which can be achieved for conv2D layers depending on the bus width and crossbar dimension.
The figure demonstrates that the speedup limit can be reached even for small bus widths (\SI{4}{\byte}) when the total number of cores is small ($\leq 32$).
Using a large bus width of \SI{32}{\byte}, up to $128$ cores can operate in parallel.
Beyond this, the speedup limit cannot be reached.
Reaching the speedup limit for sufficiently high bus widths proves that the synchronization presented in this work does not cause long waiting times.
The highest speedup that is achieved is $16\times$ for layer $5$.
The speedup of the cyclic method is slightly higher compared to the linear method because the instructions are distributed more evenly among the cores (see \cref{sec:sync_schemes}).
Thus, the linear method should be preferred due to its simple implementation.
\begin{table}[!bp]
\centering
\vspace{-0.2cm}
\caption{\label{tab:mobilenet}Excerpt from Mobilnet's conv2D Layers}
\begin{tabular}{ c c c c }
\# & kernel shape & matrix shape & input shape \\ \hline
 1 & $1\times 1\times 128\times 128$ & $128\times 128$ & $56\times 56\times 128$ \\
 2 & $1\times 1\times 128\times 256$ & $256\times 128$ & $28\times 28\times 128$ \\
 3 & $1\times 1\times 256\times 256$ & $256\times 256$ & $28\times 28\times 256$ \\
 4 & $1\times 1\times 256\times 512$ & $512\times 256$ & $14\times 14\times 256$ \\
 5 & $1\times 1\times 512\times 512$ & $512\times 512$ & $14\times 14\times 512$ \\
 6 & $1\times 1\times 512\times 1024$ & $1024\times 512$ & $7\times 7\times 512$ \\
 7 & $1\times 1\times 1024\times 1024$ & $1024\times 1024$ & $7\times 7\times 1024$ \\
\end{tabular}
\end{table}

\subsection{Bus width}
\label{sec:results_arch_exploration}
\begin{figure}[!bp]
  \centering
  \includegraphics[ width=\linewidth, keepaspectratio]{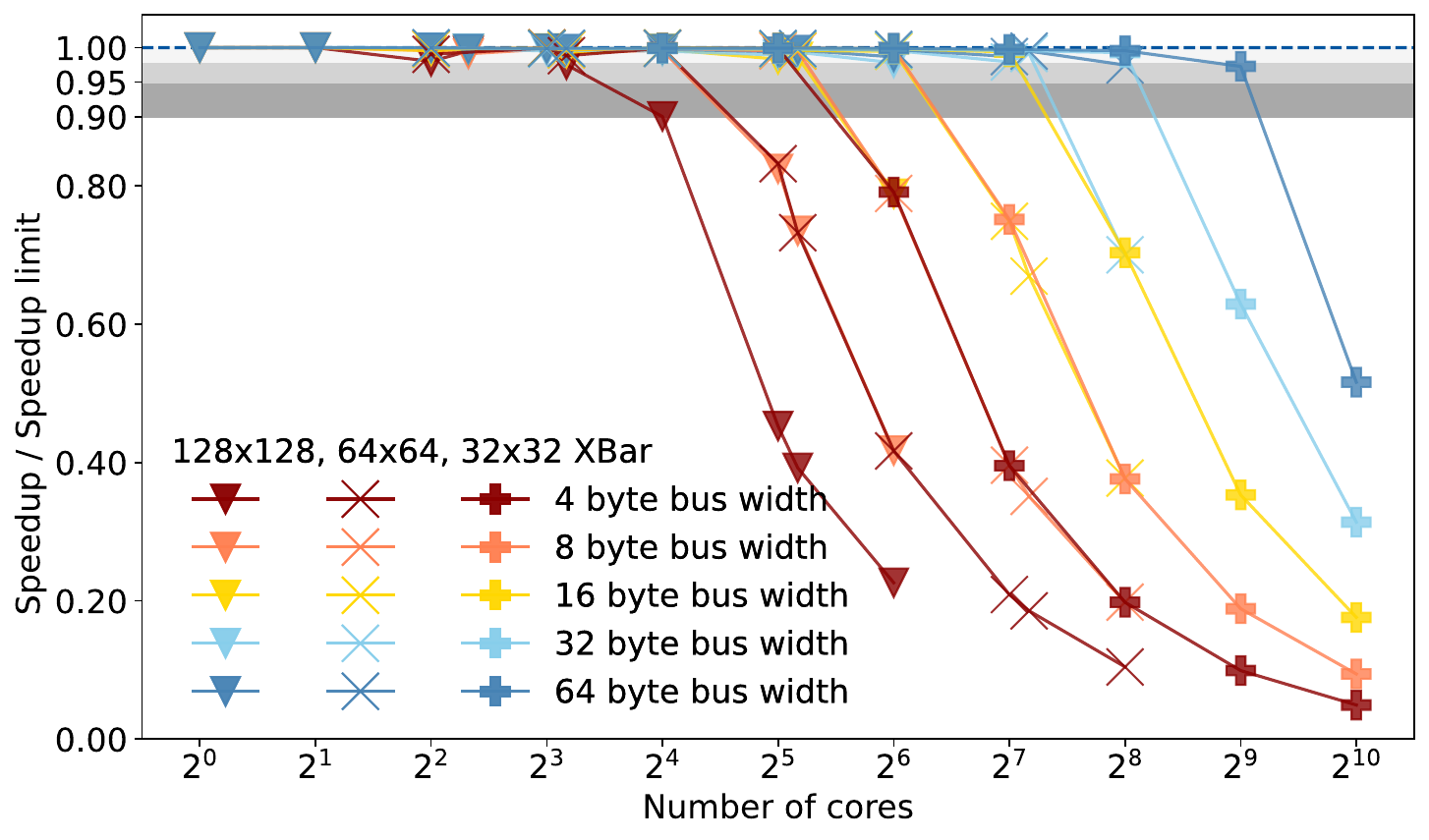}
  \caption{Speedup divided by speedup limit of cyclic synchronization scheme for different layers, bus widths, and crossbar dimensions.}
  \label{fig:results_speedup}
\end{figure}

We previously demonstrated that the synchronization methods are very efficient as the speedup limit can almost be attained.
However, this limit can only be achieved when the bus is sufficiently wide which prevents it from becoming a bottleneck.
\cref{fig:results_speedup} shows to which extent the speedup limit can be reached depending on the crossbar dimension, bus width, and the number of cores.
Each line represents one combination of crossbar dimension and bus width.
For every combination, conv2D layers from the Mobilenet and ResNet-18 architecture were compiled and simulated.
The data from \cref{fig:results_speedup} can be used to determine two things, the appropriate number of cores for a given crossbar dimension and bus width or a reasonable bus width for a given number of cores and crossbar dimension.

In general, the smaller the bus width, the lower the number of cores the bus can handle without becoming a bottleneck.
For small bus widths (\SI{4}{\byte}, red line), only a maximum of $16$ cores are worthwhile to achieve more than \SI{90}{\percent} of the speedup limit, with \SI{64}{\byte} (blue line) the system can contain up to $512$ cores.
If the crossbar dimension is halved, i.e., the number of required cores is quadrupled, then the bus width should at least be doubled to achieve similar performance.
For a given number of cores, \cref{fig:results_speedup} can be used to determine a suitable combination of crossbar dimension and bus width.

\subsection{Area overhead}
Apart from the performance, the overhead caused by synchronization needs to be considered.
A major advantage of the methods presented in this work is that synchronization can be realized by simple register accesses.
Only one register per \ac{cim} core is needed.
Assuming that one of the \SI{32}{K} attributes from \cite{ankit2019puma}
requires one byte of memory (see \Cref{sec:sync}),
our approach saves at least \SI{87.5}{\percent} of the synchronization memory,
since with a maximum of $1024$ cores and \SI{4}{\byte} per register,
only \SI{4}{\kilo\byte} of memory is required.

\subsection{Synchronization overhead}
\begin{table*}[!tb]
    \vspace{-0.3cm}
    \centering
    \caption{\label{tab:instr_load_store}Number of CALL instructions and load/store values}
    \begin{tabular}{|c|cccc|cccc|cccc|}
        \hline
        \multirow{2}{*}{\#} & \multicolumn{4}{c|}{32x32 XBar} & \multicolumn{4}{c|}{64x64 XBar} & \multicolumn{4}{c|}{128x128 XBar}\\ \cline{2-13}
        & Cores[\#] & Load{[}val{]} & Store{[}val{]} & Calls[\#]
        & Cores[\#] & Load{[}val{]} & Store{[}val{]} & Calls[\#]
        & Cores[\#] & Load{[}val{]} & Store{[}val{]} & Calls[\#] \\
        \hline
        \hline
        1 & 16 &2809856 & 1605632 & 37632& 4 &1204224 & 802816 & 6272& 1 &401408 & 401408 & 0\\
        2 & 32 &1404928 & 802816 & 18816& 8 &602112 & 401408 & 3136& 2 &200704 & 200704 & 0\\
        3 & 64 &3010560 & 1605632 & 43904& 16 &1404928 & 802816 & 9408& 4 &602112 & 401408 & 1568\\
        4 & 128 &1505280 & 802816 & 21952& 32 &702464 & 401408 & 4704& 8 &301056 & 200704 & 784\\
        5 & 256 &3110912 & 1605632 & 47040& 64 &1505280 & 802816 & 10976& 16 &702464 & 401408 & 2352\\
        6 & 512 &1555456 & 802816 & 23520& 128 &752640 & 401408 & 5488& 32 &351232 & 200704 & 1176\\
        7 & 1024 &3161088 & 1605632 & 48608& 256 &1555456 & 802816 & 11760& 64 &752640 & 401408 & 2744\\
        \hline
    \end{tabular}
\end{table*}
Synchronization requires additional operations. In contrast to the WAIT operation, the CALL operation must be transferred over the bus, which increases bus traffic.
The smaller the crossbar size, the more cores are needed, and the more CALL operations have to be executed.

\begin{figure}[!hbp]
  \centering
  \includegraphics[ width=\linewidth, keepaspectratio]{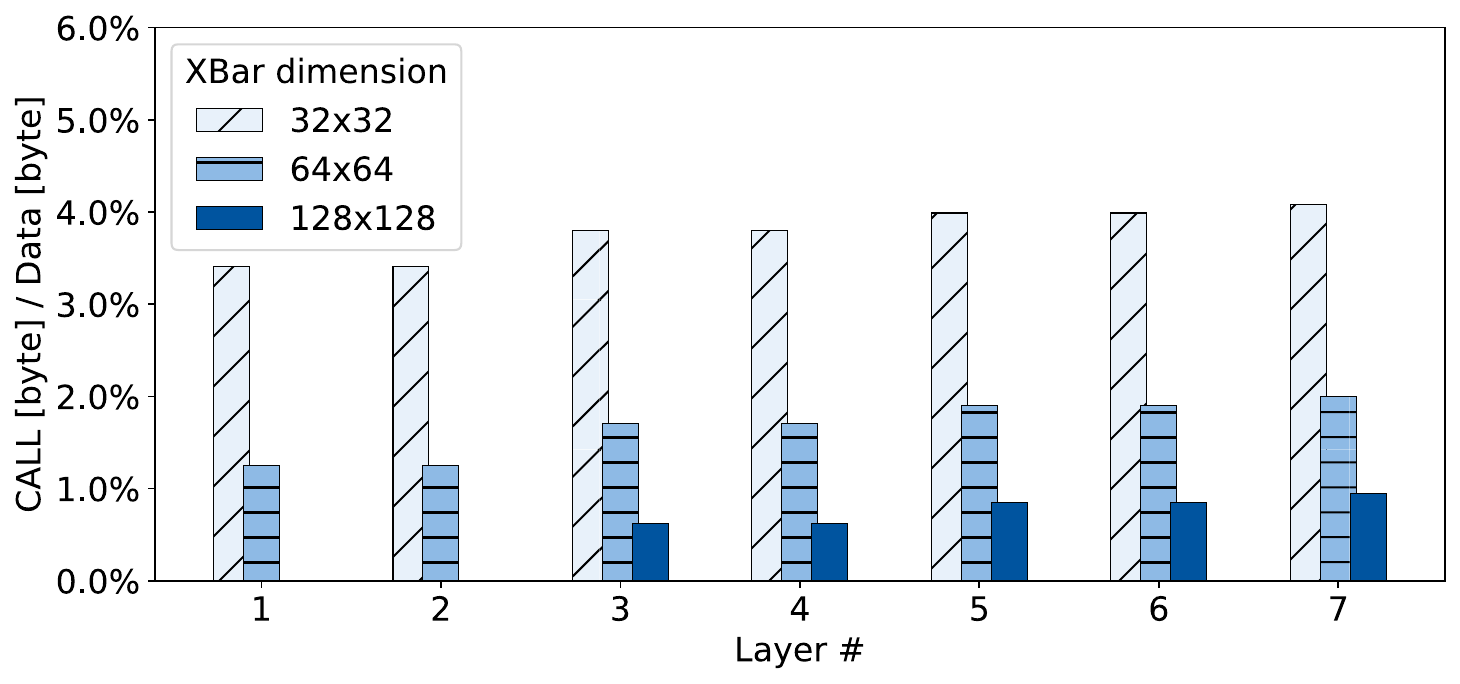}
  \caption{Bus traffic caused by CALL operations (\SI{4}{\byte} per operation) in relation to transferred data (\SI{1}{\byte} per data value).}
  \label{fig:results_speedup_perc}
\end{figure}

\cref{fig:results_speedup_perc} shows that the bus traffic caused by CALL operations is small compared to the data values transferred over the bus.
For CALL operations with a size of \SI{4}{\byte} and data values with a size of \SI{1}{\byte}, the overhead is less than \SI{4}{\percent} when using $32\times 32$ crossbars, less than \SI{2}{\percent} when using $64\times 64$ crossbars, and less than \SI{1}{\percent} when using $128\times 128$ crossbars.

\cref{tab:instr_load_store} shows the absolute number of LOAD and STORE operations.
Note that the number of loaded values is greater than the number of values in the \ac{ifm} and the number of stored values is greater than the number of values in the \ac{ofm}.
The reason for this is that the exchange of partial results and the loading of bias values are also counted.
Some input values are loaded multiple times because the same input values are multiplied by different kernel values (see~\cite{rhe2022vwc}).

\vspace{-0.5cm}
\section{Conclusion}

This paper proposes efficient, low-overhead synchronization techniques to enable the parallel execution of single layers on \ac{rram}-based multi-core \ac{cim} architectures.
We introduce an architecture that supports synchronization and data exchange in a decentralized and event-based manner.
The synchronization mechanisms require significantly less memory compared to the state of the art.
On the compiler side, we generate code for different architecture setups and evaluate them on a simulator.
By exploiting the synchronization mechanisms of the architecture,
we achieve more than \SI{99}{\percent} of the theoretical acceleration limit for conv2D layers of state-of-the-art \acp{cnn} with less than \SI{4}{\percent} additional bus traffic.

This work contributes to understanding the challenges of mapping \acp{cnn} to multi-core \ac{cim} systems. 
The presented techniques can be used as building blocks for compilers to enable parallel inference of \ac{cnn} layers.
As a future step, data dependencies between different layers must be considered to enable full system-level integration.

\bibliographystyle{IEEEtran}
\bibliography{bibtexentry}

\end{document}